\documentclass{revtex4}
\usepackage{graphicx}
\begin{document}

\title{HAWC: A Next Generation All-Sky VHE Gamma-Ray Telescope}

\author{G. Sinnis}
\email{gus@lanl.gov}
\affiliation{Los Alamos National Laboratory, Los Alamos, NM 87545}
\author{A. Smith}
\email{asmith@umdgrb.umd.edu}
\affiliation{University of Maryland, College Park, MD 20742}
\author{J. E. McEnery}
\email{mcenery@milkyway.gsfc.nasa.gov}
\affiliation{NASA Goddard Space Flight Center, Greenbelt, MD 20771}

\begin{abstract}
The study of the universe at energies above 100 GeV is a relatively new and 
exciting field.  The current generation of pointed instruments have detected 
TeV gamma rays from at least 10 sources and the next generation of detectors
promises a large increase in sensitivity.  We have also seen the development
of a new type of all-sky monitor in this energy regime based on water 
Cherenkov technology (Milagro).  To fully understand the universe at these 
extreme energies requires a highly sensitive detector capable of 
continuously monitoring the entire overhead sky.  Such an instrument 
could observe prompt emission from gamma-ray bursts and probe the 
limits of Lorentz invariance at high energies. 
With sufficient sensitivity it could detect short transients ($\sim$15 minutes)
from active galaxies and study the time structure of flares at energies
unattainable to space-based instruments.  Unlike pointed instruments a 
wide-field instrument can make an unbiased study of all active galaxies 
and enable many multi-wavelength campaigns to study these objects.  
This paper describes the
design and performance of a next generation water Cherenkov detector.  To
attain a low energy threshold and have high sensitivity the detector 
should be located at high altitude ($>$ 4km) and have a large area 
($\sim$40,000 m$^2$).  Such an instrument could detect gamma ray bursts out to
a redshift of 1, observe flares from active galaxies as short as 15 minutes 
in duration, and survey the overhead sky at a level of 50 mCrab in one
year.
\end{abstract}

\maketitle

\section{Introduction}
\label{sec:Introduction}

The past 15 years have seen large advances in the capabilities of ground-based 
gamma ray detection, from the pioneering observation of the Crab Nebula by 
the Whipple observatory in 1989\cite{weekes1989} to the new generation of 
air Cherenkov telescope arrays such as HESS\cite{hofmann2003}, 
VERITAS\cite{veritas}, and CANGAROO\cite{yoshikoski1999}  and large 
area air Cherenkov telescopes such as STACEE \cite{hanna2002}, 
CELESTE\cite{pare2002}, and MAGIC\cite{lorenz2003}.  There are 
now at least 10 known sources of very-high-energy (VHE) gamma rays\cite{horan2003}.  
The physics of these objects is astounding: from 
spinning neutron stars to super-massive black holes, these objects 
manage to accelerate particles to energies well in excess of 10 TeV.  
How this acceleration occurs is not well 
understood and there is not universal agreement on what particles are 
being accelerated in some of these sources.  At lower energies EGRET has 
detected over 270 sources of high-energy gamma rays\cite{hartman1999} 
and GLAST is expected to detect several thousand sources.  In addition 
there are transient sources such as gamma-ray bursts that have to date 
eluded conclusive detection in the VHE regime (despite some 
tantalizing hints\cite{atkins2000}).  The paucity of VHE sources 
can be traced to the nature of the existing instruments: they are 
either narrow field instruments that can only view a small region 
of the sky at any one time and can only operate on clear moonless nights 
(Whipple, HEGRA, etc.) or large field instruments with limited 
sensitivity (Milagro, Tibet Array).  
	
The Milagro observatory has pioneered the use of a large area water 
Cherenkov detector for the detection of extensive air showers.  Since an 
extensive air shower (EAS) array directly detects the particles that 
survive to ground level it can operate continuously and simultaneously 
view the entire overhead sky.  With the observation of the Crab Nebula 
and the active galaxies Mrk 421 and Mrk 501, Milagro has proven the 
efficacy of the technique and its ability to reject the cosmic-ray 
background at a level sufficient to detect sources\cite{atkins2003}.  
At the same time the Tibet group\cite{amenomori1999} has 
demonstrated the importance of a high-altitude site and what 
can be accomplished with a classical scintillator array at extreme altitudes.

A detector with the all-sky and high-duty factor capabilities of Milagro,
but with a substantially lower energy threshold and a greatly increased 
sensitivity, would dramatically improve our knowledge of the VHE universe.   
Reasonable design goals for such an instrument are:
\begin{itemize}%
   \item Ability to detect gamma-ray bursts to a redshift of 1.0
   \item Ability to detect AGN to a redshift beyond 0.3
   \item Ability to resolve AGN flares at the intensities and durations
	observed by the current generation of ACTs
   \item Ability to detect the Crab Nebula in a single transit
\end{itemize}

This paper describes a design for a next generation all-sky VHE 
gamma-ray telescope, the HAWC (High Altitude Water Cherenkov) array, 
that satisfies these requirements.  To quantify the definition of 
observing ``short'' flares from AGN, previous 
measurements of flare intensities and durations by air Cherenkov telescopes
can be used.  To date the shortest observed flares have had 
$\sim$15 minute durations with an intensity of 3-4 times 
that of the Crab\cite{gaidos1996}.  The low energy threshold needed to 
accomplish these goals requires that the detector be 
placed at extreme altitudes
(HAWC would be situated at an altitude of $\sim$4500 meters) and the required 
sensitivity demands a large area detector - of order 40,000 m$^2$.  Section 
\ref{sec:Particle_Detection} discusses the limiting performance of an EAS 
array based on the properties of the EAS, 
section \ref{sec:Detector_Description} gives a physical 
description of the HAWC and section \ref{sec:Detector_Performance} 
details the expected performance of HAWC.

\section{Particle Detection Arrays in VHE Astronomy}
\label{sec:Particle_Detection}
The ultimate performance of an EAS array
will be determined by the number, type, position, arrival time, 
and energy of the particles that reach the ground.  
Here these properties of air showers are investigated
to arrive at the limiting performance of EAS arrays.  
To attain this level of performance an EAS array would have to 
measure each of the above parameters with good precision.  

The most well-studied aspect of EAS is the dependence of the number of
particles to reach ground level on the observation altitude.  
For electromagnetic cascades, approximation B is a good estimator 
of the average number of particles in an EAS as a function
of atmospheric depth.  However, at the threshold of an EAS array, 
it is the fluctuations in the shower development that determines 
the response of the detector.  To incorporate the effect of shower 
fluctuations the event generator CORSIKA (version 6.003\cite{knapp1993}) 
is used to generate EAS from gamma rays.  The gamma rays were generated 
with an $E^{-2.4}$ spectrum beginning at 10 GeV, and 
uniformly over over the sky with zenith angles from 0 to 45 degrees.  
Four different observation altitudes were studied: 
2500m, 3500m, 4500m, and 5200m.  

\begin{figure}[ht]
\noindent
\hskip -0.3in
\includegraphics[height=0.3\textheight]{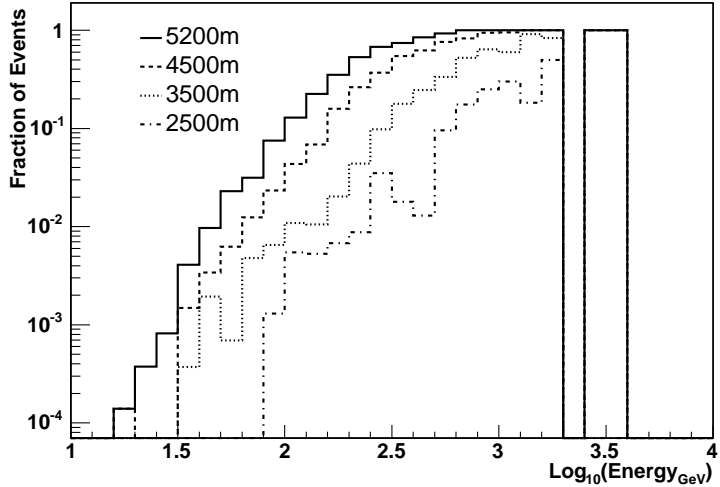}
\caption{The fraction of gamma-ray primaries that result in an EAS 
with more than 100 particles each of more than 10 MeV energy as a 
function of primary energy. The results for four different
observation levels are shown.\label{fig:f1-altitude_effect}}
\end{figure}

Figure \ref{fig:f1-altitude_effect} shows the fraction of primary 
gamma rays that generated an air shower where more than 100 particles 
with energy above 10 MeV survived to the observation level.  The 
requirement that a particle have at least 10 MeV
in energy is imposed as a reasonable detection criterion and the 
requirement that 100 such particles survive enables one to 
examine the effects of altitude on the fluctuation spectrum of the 
air showers.  This figure is a reasonable indication of the
relative effective area of a fixed detector design as a function 
of the altitude of the detector.  At high energies each km in altitude 
results in a factor of 2-3 increase in effective area.  
At low energies (of relevance for extragalactic sources such as GRBs) 
the increase with altitude is larger. Note that for primary 
energies between 100 GeV and 500 GeV a detector placed at 5200m 
has $\sim50$ times more effective area than the identical detector 
placed at 2500m altitude (close to the altitude of Milagro).  
From this figure alone one can estimate that a detector roughly 
10 times the size of Milagro placed at 5200m altitude would be
of order 22 ($\sqrt{50\times10}$) times more sensitive than Milagro 
(assuming the background scaled in a similar fashion).  
This level of sensitivity satisfies the requirements listed above.

\begin{figure}[ht]
\noindent
\hskip -0.3in
\includegraphics[height=0.3\textheight]{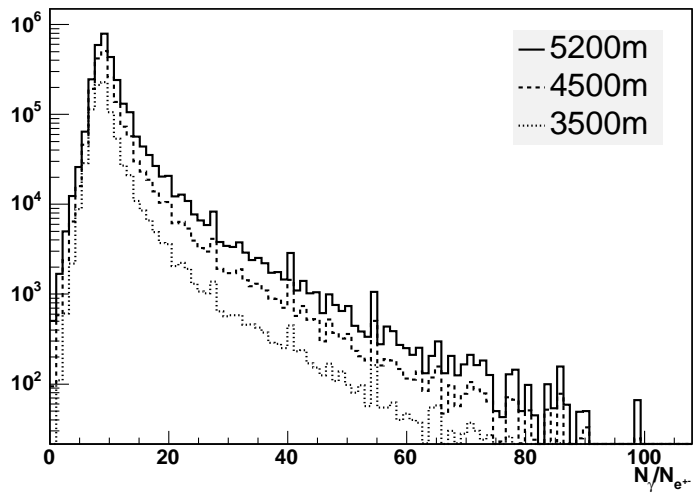}
\caption{The ratio of gamma rays to electrons and positrons in the 
extensive air shower at ground level for three different observation 
altitudes.\label{fig:gamma-electron_ratio}}
\end{figure}

It is well known that gamma rays outnumber electrons and 
positrons in an air shower by a large factor.  
Figure \ref{fig:gamma-electron_ratio} 
shows the ratio of gamma rays to electrons ($>$10 MeV) for three 
observation altitudes: 3500m, 4500m, and 5200m.  Over this altitude 
range the ratio of gamma rays to electrons is relatively independent 
of the observation level.  At these low energies gamma rays outnumber 
electrons and positrons by an order of magnitude.  Therefore 
to attain the lowest possible energy threshold the detector must be 
sensitive to the gamma rays in the air shower.  

\begin{figure}[ht]
\noindent
\hskip -0.3in
\includegraphics[height=0.3\textheight]{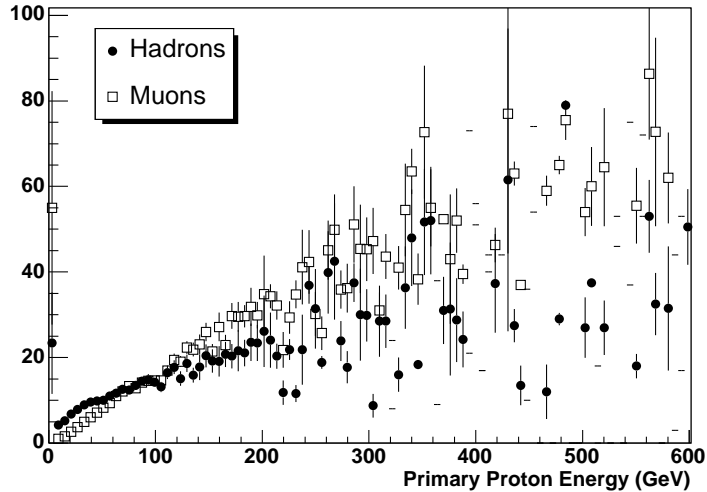}
\caption{The average number of hadrons and muons in an extensive air 
shower as a function of energy.  In this figure an observation 
altitude of 5200m was assumed.\label{fig:hadron_muon}}
\end{figure}

An important feature of any gamma-ray detector is its ability to 
reject a significant fraction of the cosmic-ray background.  
The Milagro detector has demonstrated that detecting and characterizing
the penetrating component of an air shower is a reliable method of 
rejecting the cosmic-ray background\cite{atkins2003}.  
In Figure \ref{fig:hadron_muon} we show the average number of hadrons 
and muons in an air shower as a function of primary proton energy.  
An observation altitude of 5200m is assumed in this figure.  
It should be noted that the fluctuations about these mean values are 
{\bf not} Poisson.  But it is clear from this figure that the ability to 
detect showering hadrons in the air shower is important for 
the discrimination of the cosmic-ray background.  
This is especially true at low energies where the hadronic 
component is larger than the muon component.  (Hadrons are defined 
to be protons, anti-protons, neutrons, and pions.)

\begin{figure}[ht]
\noindent
\hskip -0.3in
\includegraphics[height=0.3\textheight]{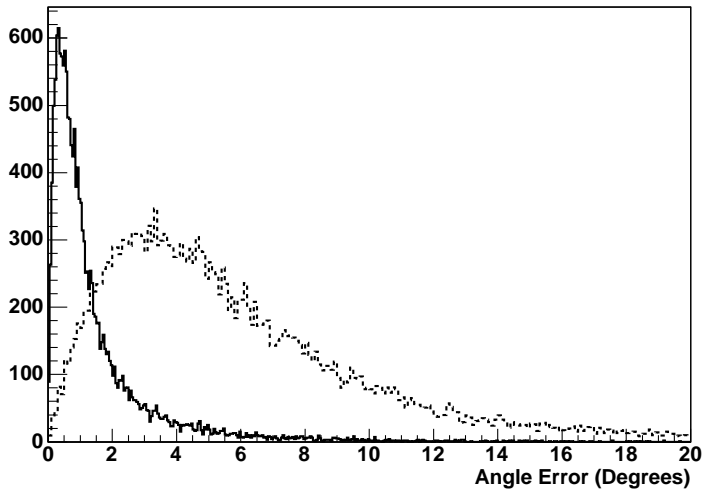}
\caption{The point spread function for the reconstruction of the 
direction of the primary particle.  The angular reconstruction is 
performed using the actual particle arrival times, positions, and energies.
The solid line shows the point spread function after an energy 
dependent curvature correction is made to the particle arrival times 
and the dashed line shows the point-spread function before
these corrections are made.\label{fig:particle-point-spread}}
\end{figure}

\begin{figure}[ht]
\noindent
\hskip -0.3in
\includegraphics[height=0.3\textheight]{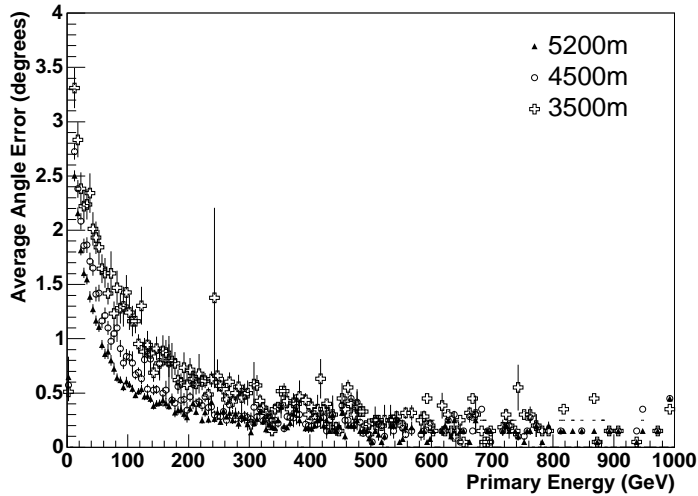}
\caption{The average angular error as a function of primary energy for
gamma-ray primaries.  The results for three different observation 
altitudes are shown. \label{fig:particle-angle_compareAltitude}}
\end{figure}

After an event is recorded its direction in the sky must be reconstructed.  
To establish a limiting angular resolution
for an EAS array gamma-ray induced EAS are generated using 
CORSIKA as above (on an $E^{-2.4}$ spectrum beginning
at 10 GeV), however only vertical showers were generated.
This simplifies the reconstruction of the air shower.  In what follows it 
is assumed that the position, time, and energy of each particle are 
measured exactly.  Since the direction of the individual 
particle is not utilized in what follows it may be possible to develop 
a detector with improved angular resolution, though at a rather large 
expense.  It is beyond the scope of this paper to give
a detailed description of the fitting procedure.  
The leading edge of the EAS is curved: particles farther from the 
core of the air shower arrive late relative to the particles at the core.  
Studies of the CORSIKA showers show that this curvature is dependent upon 
the individual energies of the shower particles.  Lower energy particles 
are delayed longer than higher energy particles.  
Before using the arrival times of the particles to fit the 
direction of the primary gamma ray the particle arrival times must be 
corrected for this effect.  To demonstrate the size of this effect 
Figure \ref{fig:particle-point-spread} shows the point spread function 
before and after this correction is applied to the particle arrival 
times (for an observation altitude of 5200m). The median angle error 
before the corrections are applied is 5.1 degrees and after the 
corrections are applied the median angle error is reduced to 
0.8 degrees (71\% of the events survive the complete fitting procedure).  
Figure \ref{fig:particle-angle_compareAltitude} shows the energy dependence
of the mean of the point-spread function as a function of primary energy.  
For a fixed primary energy the angular reconstruction is improved by 
moving the detector to higher altitude.  A detector at 4500m would 
have the same angular resolution as a detector at 5200m for gamma-ray 
primaries with 50\% more energy.  A detector at 3500m altitude would 
require over a 100\% increase in primary energy to obtain a similar
angular resolution.

\begin{figure}[ht]
\noindent
\hskip -0.3in
\includegraphics[height=0.3\textheight]{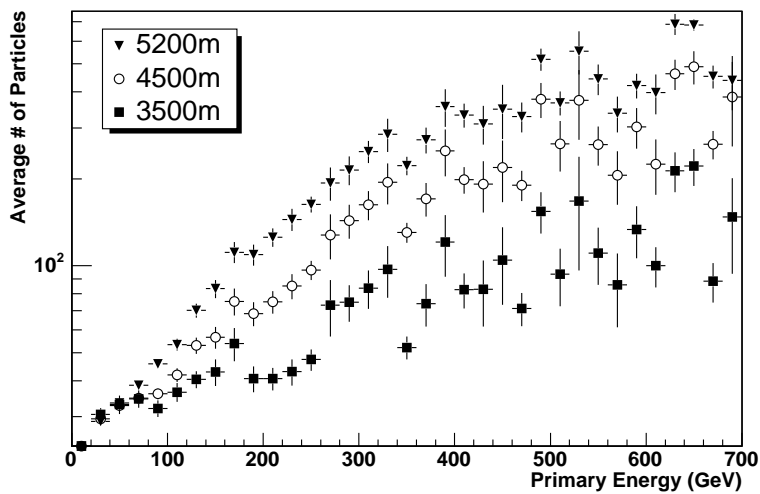}
\caption{The average number of particles with more than 10 MeV energy that 
reach the ground within 100m of the shower core as a function of primary 
energy for observation altitudes of 3500m, 4500m, and 5200m. The parent 
distributions are asymmetric so that the most probable value is somewhat 
smaller than the average value at each energy.
\label{fig:eres-particles_compare}}
\end{figure}

\begin{figure}[ht]
\noindent
\hskip -0.3in
\includegraphics[height=0.3\textheight]{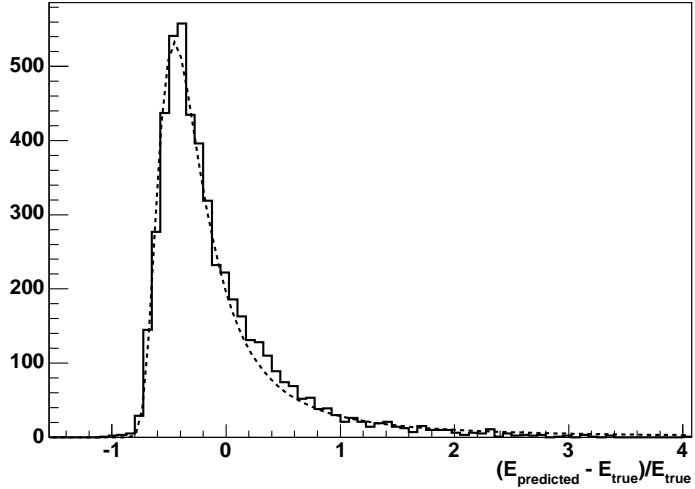}
\caption{The intrinsic energy resolution of a detector at an altitude of 5200m.
This resolution is obtained by counting the number of particles
with more than 10 MeV that fall within 100m of the shower core.  
The effect of finite detector resolution has not been accounted for in this figure.  
The fit curve is to a Landau distribution. The fit parameters are: most 
probable value is -0.42 and sigma is 0.13\label{fig:particle_eres_gt_50GeV}}
\end{figure}

Because of the fluctuations in the shower development (mainly the depth of 
the first interaction) EAS arrays have relatively poor energy resolution.   
To investigate the limiting energy resolution of an EAS array primary gamma 
rays were generated on an $E^{-2.4}$ spectrum from the zenith and the total 
number of particles with more than 10 MeV that reach the ground within 100m 
of the shower core were counted.  Figure \ref{fig:eres-particles_compare} 
shows the mean (and the error on the mean) number of particles on the ground 
as a function of primary energy for observation altitudes of 3500m, 4500m, and 
5200m.  At an altitude of 5200m there is a good correlation between 
the mean number of particles and primary energy at primary energies 
as low as 50 GeV.  At an altitude of 4500m this correlation begins 
at about 100 GeV in primary energy and at 3500m altitude there is no 
correlation until the primary energy is about 250 GeV.  To estimate 
the energy resolution, the curves in Figure \ref{fig:eres-particles_compare} 
are fit to a power law in the number of particles that reach the ground 
($E=a_0 + a_1\times N + a_2\times N^2$, where $N$ is the number of 
particles on the ground).  Figure \ref{fig:particle_eres_gt_50GeV} 
shows the distribution of $(E_{predicted}-E_{true})/E_{true}$ for 
primary energies above 50 GeV (for an observation altitude of 5200m).  
The resulting distribution is non-Gaussian, but is fit reasonably well 
by a Landau distribution (shown on the figure).  The asymmetric tail is 
an over-estimation of the particle energy and is inherent in the nature 
of the fluctuations in the depth of the first interaction.  With a full 
width at half maximum of 0.5 the equivalent size of the 1-sigma error 
bar would be $\sim$25\% (if the distribution were Gaussian).  As one 
goes to higher energy primaries the distribution becomes narrower and
more Gaussian.  At energies below 50 GeV the energy resolution is 
significantly worse, with an RMS greater than 1.  Since the fit is to 
the mean energy and there is a large positive tail, the peak in the 
distribution is negative.

This section concludes with a brief summary of the best performance that 
can be expected from an extensive air shower array operating at high 
altitude.  
\begin{itemize}
\item An energy threshold below 50 GeV
\item An angular resolution of 0.25 degrees at median energy
\item An energy resolution $\sim$25\% above 50 GeV (with a non-Gaussian tail)
\end{itemize}

\section{Detector Description}
\label{sec:Detector_Description}
Based on the above study and experience with the Milagro detector 
a complete Monte Carlo simulation of a detector that is similar to Milagro
has been performed.  The need to detect the gamma rays in the EAS and the 
penetrating component of muons and hadronic
cascades leads to a thick active detector.  Water is a natural and 
inexpensive choice of detecting medium.  The Cherenkov photons are 
relatively plentiful, prompt, and have large mean free paths (relative 
to the water depth) in clean water.  Since the Cherenkov angle in water 
is 41$^{\circ}$, a detector spacing of roughly twice the detector 
depth ensures that there are no geometrical blind areas.  (The depth 
and spacing were chosen to match that of Milagro,
though more work is needed to optimize these parameters.)
The choice of a two layer design was again driven by 
experience with Milagro.  The deep layer makes a good calorimeter, whereas a 
shallow layer does not.  In the shallow layer the detected light level
depends critically on the exact geometry of the incoming particles, their 
distance from the photomultiplier tube (PMT) and incident angle.  These 
fluctuations caused by the geometry of the incoming
particles dilutes the energy measuring capability of a shallow layer of PMTs.  
The ability to measure the energy of the shower particles is needed to 
determine the penetrating component of the air shower.  In Milagro this 
is the most useful method of rejecting the abundant cosmic-ray background.
A shallow layer may be required to obtain the best possible angular resolution 
(also important for rejecting the cosmic-ray background).  
Since the speed of light in water is substantially different
from the speed at which the particles propagate (the speed 
of light in vacuum), the light and the particles diverge.  
This leads to an increase in the  dispersion of the arrival time 
of the measured shower front as the depth of the measuring 
device is increased and the angular resolution is proportional 
to this spread.  This effect is somewhat compensated for by the 
fact that a deeper layer typically makes more measurements of 
the shower front (because each PMT can see a larger 
area of the water surface).  The top layer may also be useful in the rejection 
of the cosmic ray background (described below).  Work in better understanding 
these effects is still underway, but it may be possible to 
eliminate the shallow layer and maintain the performance of the detector.

In the remainder of this paper, the following detector 
characteristics are used.
\begin{itemize}%
   \item 40,000m$^2$ physical area (200m$\times$200m)
   \item 2.7m spacing of PMTs (74$\times$74 PMTs per layer) 
   \item Hamamatsu 20cm PMTs (R5912SEL)
   \item Two layers of PMTs. A shallow layer under 1.5m of water 
and a deep layer under 6.5m of water
   \item An attenuation length of 20 meters, composed solely of absorption.
   \item Trigger requires 50 PMTs in the top layer to be struck by at 
least 1 photo-electron (PE)
   \item A detector latitude of 36 degree (that of Milagro)
\end{itemize}
CORSIKA 6.003\cite{knapp1993} was used to generate the air showers and 
GEANT 3.21 was used to track the particles through the water, generate 
the Cherenkov light in the water, and to model the detector response 
(PMTs, reflectivity of the material, scattering and absorption of 
light in water, etc.). For the remainder of this paper this design 
will be referred to as the High Altitude Water Cherenkov (HAWC) array.

\section{Detector Performance} 
\label{sec:Detector_Performance}

\subsection{Event Reconstruction}
\label{sec:Event_Reconstruction}

To obtain realistic estimates of the sensitivity of HAWC, the individual
events must be reconstructed - their core position, direction, type (hadronic
or gamma ray), and energy need to be determined.   Preliminary algorithms
have been developed to determine all of these parameters with the 
exception of the primary energy.  Work is ongoing in improving the 
existing algorithms and developing one to determine the energy of 
the primary gamma ray.  In what follows, the events were generated 
beginning at 10 GeV on an E$^{-2.49}$ spectrum (similar to that 
observed for the Crab Nebula\cite{weekes1989}) uniformly over 
incident direction.

\begin{figure}[ht]
\noindent
\hskip -0.3in
\includegraphics[height=0.3\textheight]{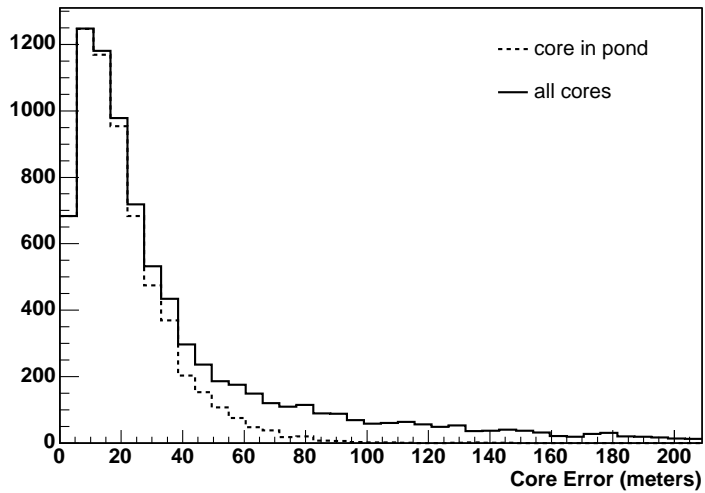}
\caption{The core error in meters using a center-of-mass core algorithm.  
Gamma ray events with 50 PMTs in the top layer, at least 20 PMTs used 
in the angular fit are selected.  The solid line shows all events and the 
dashed line contains only events whose core fell inside the array.
\label{fig:hawc-core_resolution4572}}
\end{figure}

As discussed above, the shower front of an EAS is not a true plane.  
Therefore, the first step in performing an angular reconstruction 
is the determination of the core of the air shower.   
A simple center-of-mass algorithm using the pulse heights recorded 
in the PMTs in the bottom layer of the detector is used to determine
the shower core.  Figure \ref{fig:hawc-core_resolution4572} shows 
the distribution of core reconstruction errors in meters.  For all 
events that trigger the detector and are successfully reconstructed 
(see below) the mean core error is 36 meters.  Despite the simplicity 
of the algorithm the core resolution is good enough that it does not 
degrade the angular resolution.  The events with large core errors are 
those whose cores fall outside of the detector.  The dashed line in the 
figure shows the core error for events whose core falls within the detector, 
the mean of this distribution is 20 meters.

\begin{figure}[ht]
\noindent
\hskip -0.3in
\includegraphics[height=0.3\textheight]{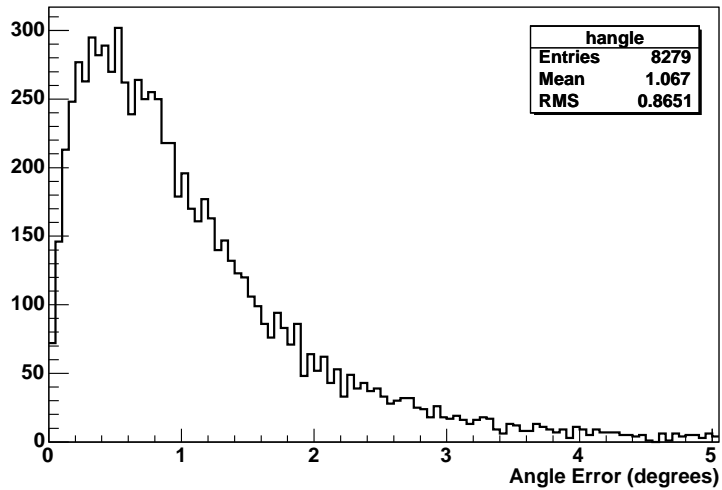}
\caption{The space angle difference distribution between the true direction
and the fit direction of gamma ray events.  This figure includes
all gamma rays that triggered the detector and had more than 20 PMTs survive
the fitting procedure. \label{fig:hawc-angular_resolution4572}}
\end{figure}

The angular reconstruction is performed in an analogous manner to that described above,
however the timing correction that was applied based on the individual particle
energies is replaced by a correction based upon the number of PEs measured in each 
PMT.  After applying this correction and a correction for the shower front
curvature (proportional to the distance from the PMT to the core of the shower)
the resultant times and positions to a plane.  
In addition the weight associated with each PMT is determined by the measured pulse height.  
Though the weight also depends upon the distance to the shower core, this effect has been 
ignored in the following analysis.  A 20-30\% improvement
in the angular resolution may be expected with an improved algorithm.  
This improvement should arise when the correlations between the two corrections 
(core distance and pulse height) are utilized and the weights assigned to each PMT 
incorporate the core distance. The fitting procedure is iterative, where the PMTs 
that made large contributions to the chi-square are removed in subsequent iterations.   
Events with more than 20 PMTs surviving in the final iteration of the fitting procedure 
are considered to be successfully fit.  The space angle difference between the true 
direction and the fit direction is shown in Figure \ref{fig:hawc-angular_resolution4572}.  
Using this point-spread function the signal to noise ratio is maximal for an 
analysis bin size of 1.2 degrees.  If the point-spread function were Gaussian this
bin size would correspond to an angular resolution of 0.75 degrees\cite{alexandreas1993}.

\begin{figure}[ht]
\noindent
\hskip -0.3in
\includegraphics[height=0.75\textheight]{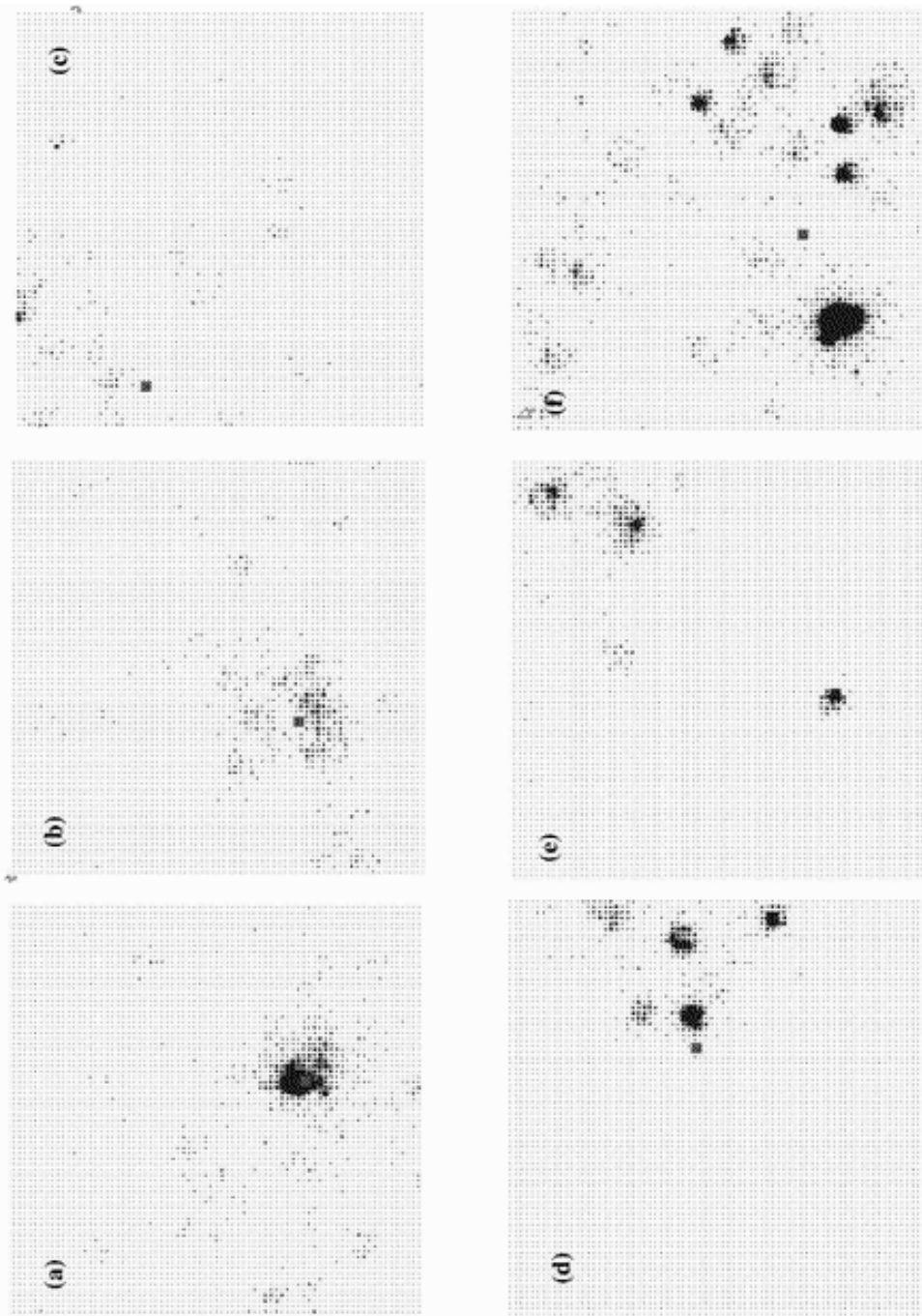}
\caption{Three gamma-ray induced events (top) and three proton induced events (bottom)
as observed in the bottom layer of HAWC. 
The small black dots represent the positions of all the PMTs.  The area of each PMT 
is proportional to the detected pulse height in the PMT and the square is the location of 
the shower core.  The energies of the primary particles are as follows: (a) 30 GeV (b) 70 GeV
(c) 230 GeV (d) 20 GeV (e) 70 GeV and (f) 270 GeV.  \label{fig:events}}
\end{figure}

The ability to discriminate between gamma-ray induced air showers and hadronic air showers 
is critical to the success of any gamma-ray detector.   Unlike air Cherenkov telescopes 
the background rejection discussed here is essentially independent of angular resolution.  
Based on angular resolution alone with a 2sr field-of-view HAWC (and Milagro) remove roughly 
99.93\% of the background in their field-of-view.  The background rejection discussed below 
is based purely on differences in the air showers regardless of their incoming direction.  
Figure \ref{fig:events} shows three gamma-ray induced events and three proton induced
events in HAWC.  The figure shows the pulse height distribution of the PMTs in the bottom layer
of the detector.  The area of each square is proportional to the number of detected 
photoelectrons (PEs) in the PMT.  The small black dots represent the PMTs and the 
square is the position of the shower core.  The figure caption gives the primary 
energy for each event.  There are clear differences in morphology between proton 
and gamma-ray events.  Though there are some proton induced events that have a 
similar morphology to gamma-rays induced events, the events shown here are 
typical in the sense that most events have similar properties.  Note that while 
some of these events have very low primary energies, they are easily detected in 
HAWC, striking many PMTs.

\begin{figure}[ht]
\noindent
\hskip -0.3in
\includegraphics[height=0.3\textheight]{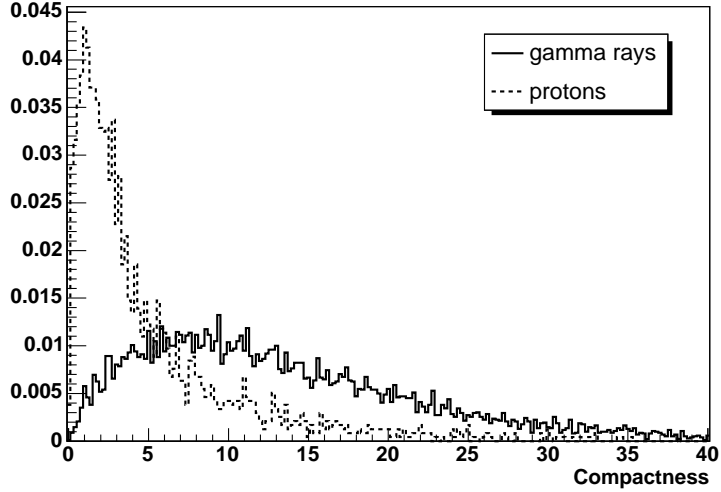}
\caption{The compactness distribution for proton induced events
and gamma-ray induced events. \label{fig:hawc-compactness4572}}
\end{figure}

Milagro utilizes a parameter known as compactness that is defined as the ratio of the 
number of PMTs in the bottom layer with more than 2 PEs detected to the maximum number 
of PEs measured in a single PMT in the bottom
layer\cite{atkins2003}.  Hadronic showers tend to have small values of compactness 
and gamma ray showers a large value of compactness.  The studies performed here 
indicate that as one moves to higher elevations the ability of this parameter to 
discriminate between hadronic and gamma-ray initiated cascades
degrades.  This is consistent with the observations of the Milagro collaboration, who found
that the background rejection capabilities of compactness improved as the zenith angle of the
primary particle increased\cite{atkins2003}.  However, a modified
form for the compactness parameter, where the number of PMTs with more than 2 PEs is replaced
with the number of PMTs that are struck in the bottom layer, has a similar
background rejection capability to that obtained by Milagro.  Figure 
\ref{fig:hawc-compactness4572} shows this modified compactness parameter 
(which will be referred to simply as compactness in what follows) for proton 
(dashed line) and gamma ray (solid line) showers.  If events are required to have 
a compactness greater than 8.3 one retains 66\% of the gamma ray
events and only 17\% of the proton events, yielding a quality factor of 1.6.  (The quality
factor is defined as the fraction of gamma-ray events that survive the cut divided by the
square root of the number of proton events that survive the cut, and is the relative improvement 
in sensitivity of the detector.)

\begin{figure}[ht]
\noindent
\hskip -0.3in
\includegraphics[height=0.3\textheight]{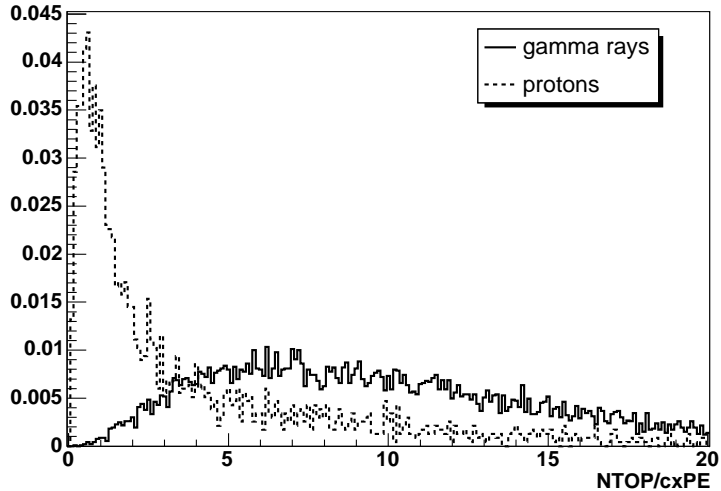}
\caption{The distribution of $NTOP/cxPE$ for proton induced events
and gamma-ray induced events. See the text for the definitions of the
parameters. \label{fig:hawc-nxTop}}
\end{figure}

Another parameter that appears promising in discriminating gamma rays from background
is the ratio of the number of PMTs struck in the top layer ($NTOP$) to the number of PEs
in the brightest PMT in the bottom layer, where the PMTs within 20m of the fit 
shower core are excluded from the
search for the maximum ($cxPE$).  This distribution is shown in Figure \ref{fig:hawc-nxTop} for
gamma ray and proton induced events.  The requirement $NTOP/cxPE >2.0$ retains
97.7\% of the gamma ray events while rejecting 52.5\% of the proton induced events.  The 
important feature of this parameter is that essentially all of the gamma ray events survive, 
so even though the quality factor is only 1.4 one has retained all of the signal events 
(regardless of their energy).  The requirement $NTOP/cxPE > 4.3$ yields a quality factor 
of 1.6 (retaining 85\% of the gamma ray induced events and 29\% of the proton induced events), 
similar to the compactness criterion, but retaining a larger fraction of the gamma ray events.  
Unlike the compactness parameter defined above (and that used by Milagro) there is
very little energy dependence to this criterion.  In Milagro compactness 
rejects low-energy gamma rays\cite{atkins2003}.

\subsection{Detector Sensitivity}
\label{sec:Detector_Sensitivity}

The performance of any detector is determined by its effective area as
a function of energy, its angular resolution, the level of the background and 
the ability to eliminate this background.  These considerations
can be parameterized in the following function:
\begin{eqnarray}
S \propto \frac{\int_{0}^{T}{\sigma(E,\theta){E^{-{\alpha}} A_{\gamma}(E,\theta(t)) dEdt}}}
	{\sqrt{\int_{0}^{T}{E^{-2.7}A_{cr}(E,\theta(t))dEdt}}}
\label{eqn:sensitivity}
\end{eqnarray}
where $S$ is the ``sensitivity'' of the detector, $\sigma(E,\theta)$ is the angular 
resolution of the detector as a function of energy and zenith angle, $\alpha$ is the 
differential spectral index of the source, $A_{\gamma}(E,\theta)$ is the effective 
area of the detector to gamma rays as a function of energy and zenith angle, $\theta(t)$ 
is the zenith angle of the source as a function of time, $A_{cr}(E,\theta)$ is the effective 
area of the detector to the cosmic-ray background as a function of energy and zenith angle, 
and $T$ is the observation time.  In determining the functions $A(E,\theta)$ the background 
rejection should be included.  The time dependence of the zenith angle is required since 
the energy response of the detector is dependent upon the atmospheric overburden.  

\begin{figure}[ht]
\noindent
\hskip -0.3in
\includegraphics[height=0.3\textheight]{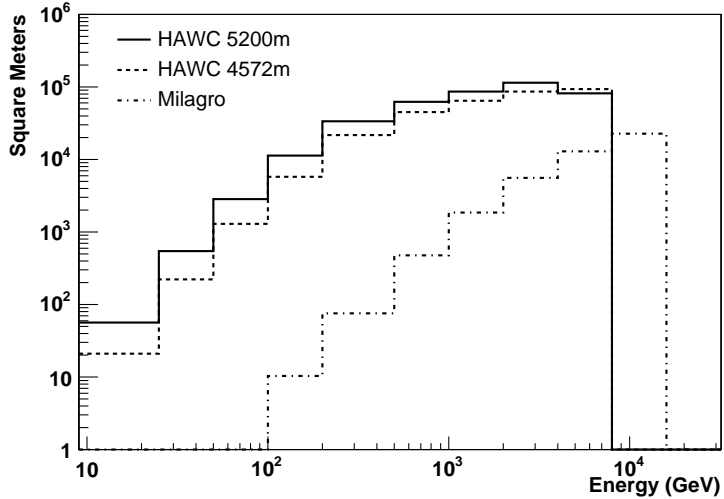}
\caption{The effective area of HAWC (to gamma rays) at 5200m and 4572m as a function of energy.
For comparison the effective area of the Milagro detector is also shown in the figure.
In calculating the effective area the following requirements were placed
on the events: $>$50 PMTs struck in the top layer, $>20$ PMTs used in the angular
reconstruction, and the reconstructed angle 
within 1.2 degrees of the true direction.\label{fig:hawc-effective_area}}
\end{figure}

To obtain the sensitivity of HAWC, equation \ref{eqn:sensitivity} is evaluated with the aid of
the Monte Carlo.  Events are generated on an E$^{-2.49}$ spectrum for gamma ray 
primaries and $E^{-2.7}$ for proton primaries.  After following
the particles and the Cherenkov light through the detector a list of PMTs that are struck, 
the number of PEs in each of these PMTs, and the arrival time of the first PE in each PMT 
is written to a file.  The events are generated over a distribution of
zenith angles to mimic an isotropic flux.  
For a given source declination the time spent at each zenith angle is found and the 
effective area as a function of energy for that zenith angle is integrated to give a number of
detected events as a function zenith angle for a given source declination.  Figure
\ref{fig:hawc-effective_area} shows the effective area as a function of energy for gamma-ray 
primaries for HAWC at two different detector altitudes, 4572m asl and 5200m asl.  
The zenith angle averaging in this figure is as if the source was spread uniformly
over the sky between zenith angles of 0 and 45 degrees.    
This figure is used for illustration only.  To calculate the sensitivity to a source
the integral in equation \ref{eqn:sensitivity} is evaluated over the source transit 
(given the declination of the source).  In the remainder of the paper (with the exception 
of the calculation of the sensitivity to gamma-ray bursts) it is assumed that the source 
is at the declination of the Crab (22 degrees) and the detector is at a latitude of 36 
degrees north.  Only events that are reconstructed within 1.2 degrees of their true 
direction are included in the calculation of the effective area.  (This accounts for 
the first term $\sigma(E,\theta)$ in equation \ref{eqn:sensitivity}.)  At high energies 
($\sim$1 TeV) HAWC at 5200m would have about 170 times the effective area of Milagro to 
gamma rays and at 100 GeV about 1000 times the effective area of Milagro.  As the energy 
decreases below this, the ratio of effective areas continues to increase, 
though there are insufficient Monte Carlo events at low energies for Milagro to determine 
the ratio below 100 GeV.  

Not withstanding the above discussion, the ``sensitivity'' of a detector is a poorly 
defined concept.  Since different astrophysical sources have 
different properties (energy spectrum, duration, prior knowledge of their existence) a single 
number called sensitivity is not useful.  In what follows the sensitivity of 
HAWC to three distinct phenomena is estimated: galactic sources with an energy 
spectrum of $E^{-2.49}$, active galaxies at various redshifts (to investigate 
the effect of the absorption of VHE photons by the extragalactic background light), 
and gamma ray bursts.  The latter is a major motivation for this type of detector, 
since ACTs by their nature have no opportunity to the prompt phase of GRB emission.

\subsubsection{Galactic Sources and Sky Survey}
The Crab Nebula is the prototypical galactic TeV gamma ray source.  In reality 
the spectrum of the Crab Nebula is not well represented by a power law over the energy
range 10 GeV to 10 TeV.  However, for simplicity and to enable a comparison to other
instruments the spectrum of the Crab Nebula\cite{hillas1998} is assumed to be 
$dN/dE=3.2\times 10^{-7}E^{-2.4}$ m$^{-2}$s$^{-1}$ 
over the energy range of HAWC.  Using the procedure outlined above the number of detected 
gamma rays per source transit in HAWC is calculated.  To estimate the background level in 
a bin of the same size with the same transit, the identical procedure is carried 
out with proton primaries (the criterion that the events must be reconstructed with 
1.2 degrees of their true direction is dropped).  The identical procedure is than performed
for the Milagro Monte Carlo. The ratio of the simulated events observed in HAWC 
to the simulated events observed in Milagro is used to scale the
actual number of events detected in Milagro and predict the background rate in HAWC.
For an observation altitude of 5200m the background rate will be $\sim$85 times that of Milagro
and for an observation altitude of 4572m the background rate will be $\sim$60 times 
larger than Milagro's.  These numbers apply to the raw data, before the application 
of any background rejection.  Table \ref{tab:crab_rates} gives the expected number 
of source and background events for a single
transit of the Crab.  The predicted excess for Milagro is somewhat larger than is actually
observed by Milagro.  With two years of data Milagro observes roughly 4$\sigma/\sqrt{\# years}$
on the Crab Nebula.  The discrepancy is accounted for by three detector related effects 
(the dead-time in Milagro is about 12\%, the on time is about 90\%, and at any given time about 6\% 
of the PMTs are not working) and one astrophysical effect, the spectrum of the Crab 
Nebula turns over at the higher energies\cite{hillas1998}.   
When these effects are properly accounted for the measurement of the Crab 
flux by Milagro is in good agreement with the measurement by HEGRA\cite{atkins2003}.  

Table \ref{tab:crab_rates} shows that a detector at 5200m elevation is about 40\% more 
sensitive than the same detector at 4572m elevation.  This is in agreement 
with the results given above on the effect of altitude on the particles at the ground.  
But both elevations result in a detector that can observe the Crab Nebula at high 
significance in a single transit.  This is an important
feature which allows one to quickly verify all aspects of the detector response.  
Current ACTs such as the Whipple telescope obtain roughly 5$\sigma$ on the Crab Nebula 
in one hour of observation.  Though the calculated transit here consists of 6 hours, the 
bulk of the signal arrives in a 4-hour span, so the sensitivity for the same observation
period of HAWC will be roughly 1/2 that of the Whipple telescope.  With one year of 
observation a 50 mCrab source would yield a 5$\sigma$ detection.  In contrast the 
VERITAS 7-telescope array will detect a 7 mCrab source at 5$\sigma$ with a 50 hour 
observation\cite{veritas}.  With this sensitivity level VERITAS would require 3-4 years 
of dedicated time to survey 2sr of the sky at the level of 50 mCrab.
(This assumes no change in sensitivity for VERITAS for sources 1 degree off-axis.)  
It is worth noting that these two surveys are fundamentally different, in that 
HAWC would obtain a measurement (or upper limit) which is an average over a year, 
while the VERITAS measurement (or upper limit) would result from a 7-minute snapshot 
over a 3-4 year period.  Given the transient nature of many VHE sources, such a snapshot 
may be a limited value.  The gamma-ray rate (in HAWC) from the Crab will be roughly 
0.4 Hz or 24/minute for a detector at 5200m and 0.23Hz (or 14/minute) for a detector 
at 4572m asl.  This number compares favorably with VERITAS and is significantly 
higher than that currently observed by the Whipple telescope.
\begin{table}[ht]
\caption{Expected Signals from the Crab Nebula in HAWC at 5200m and 4572m elevation.
Milagro is at an elevation of 2700m and is $\sim 1/10$ the size of HAWC.}
{\footnotesize
\begin{tabular}{cllllll}
\hline
{} &{} &{} &{} &{} & {} & {}\\[-1.5ex]
Elevation & Signal (Raw) & Bkgnd (Raw) & $\sigma$ (Raw) & Signal (Cut) & Bkgnd (Cut) & $\sigma$ (Cut)\\[1ex]
\hline
{} &{} &{} &{} &{} & {} & {}\\[-1.5ex]
5200m    & 5980 & 1.7e6 & 4.6  & 5030 & 6.5e5 & 6.0  \\[1ex]
4572m    & 3350 & 1.2e6 & 3.0  & 2920 & 4.5e5 & 4.3  \\[1ex]
Milagro  & 36   & 2.1e4 & 0.24 & 17   & 2300  & 0.35 \\[1ex]
\hline
\end{tabular}\label{tab:crab_rates} }
\end{table}

\subsubsection{Extragalactic Sources}

\begin{figure}[ht]
\noindent
\hskip -0.3in
\includegraphics[height=0.3\textheight]{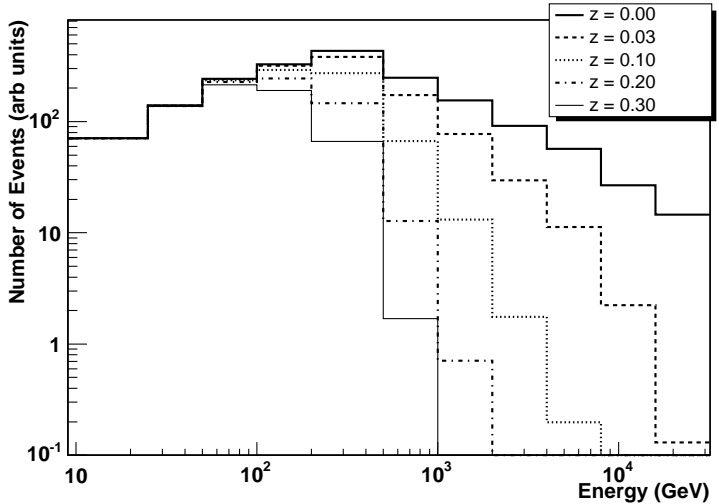}
\caption{The estimated number of detected events in HAWC as a function of energy for a source with a
differential spectral index of -2.49 placed at several different redshifts. An observation
altitude of 5200m has been assumed in this figure.  Only events that survive the background
rejection cut $nTop/cxPE$ are used to calculate the curves in the figure.\label{fig:hawc-ebl-energy}}
\end{figure}

For more distant sources one must account for the absorption of high-energy photons 
caused by interactions with the extragalactic background light (EBL)\cite{primack,stecker}.  
While a direct measurement of the EBL has proved elusive, several models exist.  In what 
follows the ``fast evolution'' model of Stecker and De Jager\cite{stecker} is used to 
estimate the effect as a function the redshift of the source on the sensitivity of HAWC.  
For nearby objects (redshift less than 0.05) there is a negligible difference between 
this model and the ``baseline'' model discussed in the reference.  At a redshift of 
0.2 the baseline model for the EBL results in $\sim$15\% more signal events detected 
by HAWC than the model used here.  Figure \ref{fig:hawc-ebl-energy} shows the number 
of signal events detected as a function of energy for sources at various redshifts.  
Only events that satisfied the background rejection criterion, $nTop/cxPE > 4.3$, 
had more the 20 PMTs survive the angular reconstruction, and were reconstructed 
within 1.2 degrees of their true direction where used to calculate the curves in the figure. 
A source differential spectral index of -2.49 and an observation altitude of 5200m have been 
assumed in this figure.  An integration of these curves gives the total number of 
detected events for each source.

\begin{figure}[ht]
\noindent
\hskip -0.3in
\includegraphics[height=0.3\textheight]{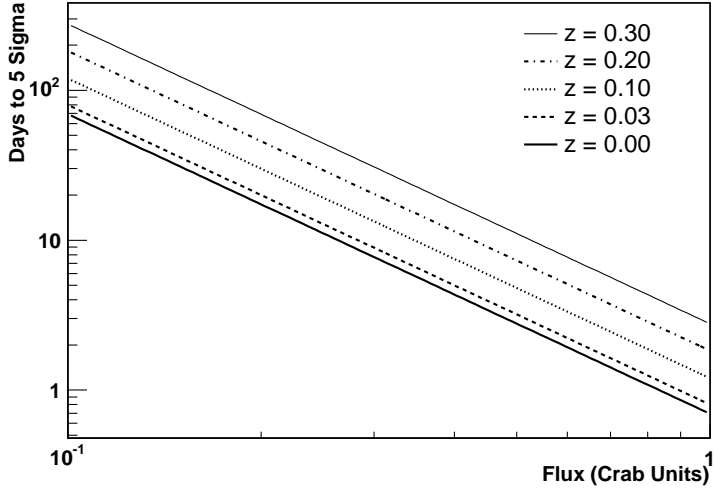}
\caption{The number of days required to detect a source at 5$\sigma$ as a function of the
source intensity for sources at various redshifts. \label{fig:hawc-agn-time5}}
\end{figure}

Using the modified spectra given in Figure \ref{fig:hawc-ebl-energy} the sensitivity of 
HAWC to distant sources can be calculated.   Figure \ref{fig:hawc-agn-time5} shows the 
number of days required to detect a source at the 5$\sigma$ level as a function of 
source intensity for several different source redshifts.  The source intensity is 
given in units of the Crab flux and is the flux that would be present at the top of 
the atmosphere if there was no absorption of the high-energy gamma rays.  HAWC could detect
a source at a redshift of 0.3 with a flux at the top of the atmosphere (before 
accounting for absorption) one tenth that of the Crab Nebula with less than one 
year of observation.

\subsubsection{Gamma Ray Bursts}
\label{sec:Gamma_Ray_Bursts}

The study of the high-energy spectrum of gamma-ray bursts (GRBs) is
perhaps the strongest motivation for an all-sky VHE telescope.
While GLAST will have good sensitivity to GRBs up to several 10's
of GeV, a ground-based instrument, with its much larger effective
area, could have good sensitivity up to the highest energy gamma rays
that reach the earth.  The ability to study the prompt phase
of VHE emission could lead to exciting new physics.
Given its low energy threshold, HAWC would have sensitivity to GRBs
at a redshift 1.

\begin{figure}[ht]
\noindent
\hskip -0.3in
\includegraphics[height=0.3\textheight]{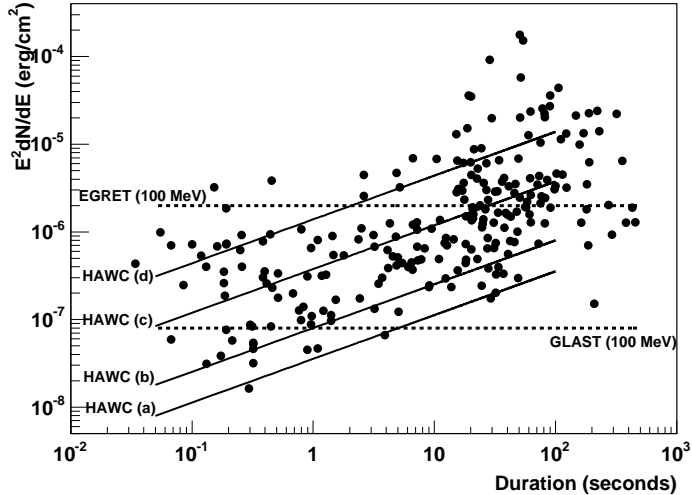}
\caption{The sensitivity of HAWC to gamma-ray bursts.  
The dots are data points from BATSE observations of burst fluence
as a function of burst duration.  The lines show the required fluence to
yield a 5$\sigma$ detection in HAWC for various cutoff energies 
(as given in the legend).  For comparison the sensitivities of
EGRET and GLAST are shown.\label{fig:hawc_grb}}
\end{figure}

To estimate the sensitivity of HAWC to GRBs requires knowledge of
any absorption at the source and the inevitable absorption of the
VHE photons as they traverse the cosmos.  Since one does not know the
inherent spectrum of GRBs, in what follows the GRB energy spectrum
is modeled as a power law with a sharp cutoff.  This cutoff could
arise from absorption at the source or from the EBL.  By investigating
the sensitivity of HAWC to different cutoff energies it is possible
to estimate the sensitivity to GRBs at different redshifts.
Figure \ref{fig:hawc_grb} shows the sensitivity of HAWC (at
an observation altitude of 4572m asl) to GRBs as a function of burst
duration for several energy cutoffs.  The spectra are assumed to be 
$E^{-2.4}$ up to the indicated cutoff energy, where the photon
flux falls to zero.  These curves give the flux level needed to
observe a GRB at the 5$\sigma$ level.  No background rejection
has been applied to these data.  The black circles are data from 
the BATSE detector, giving the observed distribution of GRB fluence
as a function of burst duration.  The lower solid line is the sensitivity of
GLAST and the upper solid line the sensitivity of EGRET.  
To detect a 1-second duration burst HAWC would require 50 events from a GRB 
over a background level of 85 events.  The signal to noise level in a detected
burst would be quite high.  

Figure \ref{fig:hawc_grb} can be used in conjunction with the
known redshift distribution of GRBs and the known GRB rate to estimate
how many GRBs/year HAWC would detect.  Assuming that the intrinsic GRB spectrum
extends to at least 50 GeV, one sees that roughly 20\% of all GRB's are
bright enough to be detected by HAWC.  If half of all bursts lie within a 
redshift of 1 and assuming a GRB rate of 1 per day, HAWC (which views 1/6 of the
sky) would detect at least 6 GRBs per year.  If the intrinsic GRB spectra
extend beyond 50 GeV HAWC would detect more bursts.

\section{Conclusions}
\label{sec:Conclusions}

A design for the next generation all-sky VHE gamma-ray telescope has
been presented.  This instrument, HAWC, would be over 20 times
more sensitive than the Milagro detector.  
With the ability to continuously view the
entire overhead sky HAWC will be an excellent complement to both GLAST and
the coming generation of air Cherenkov telescopes (HESS, MAGIC, VERITAS,
and CANGAROO III).  With comparable sensitivity to GLAST it will be the only
instrument capable of monitoring the many thousands of sources that GLAST is
expected to detect at higher energies.  In addition to searching the sky
for galactic sources (the VHE complement to the 150 EGRET unidentified objects),
and active galaxies, the low-energy sensitivity of a detector placed at high 
altitude ensures that such an instrument will detect any VHE emission from 
gamma-ray bursts.  Perhaps most importantly an open aperture instrument
with this level of sensitivity could discover completely new and unexpected
phenomena that have so far eluded detection.

\end{document}